%% file: writeup-2024-04-09.tex
\documentclass[11pt]{article} 
\usepackage{mystyle-new}
\usepackage{authblk}
\usepackage[T1]{fontenc}
\usepackage{epsfig,amsmath} 
\usepackage{hepnames,hepunits}
\usepackage{hyperref}
\usepackage{color}
\usepackage{graphicx}
\usepackage{orcidlink}
\definecolor{red}{rgb}{1,0,0}
\def\lesssim{\ \hbox{\raise 2pt \hbox{$<$} \kern -13pt
                     \lower 3pt \hbox{$\sim$}}\ }
\def\greatersim{\ \hbox{\raise 2pt \hbox{$>$} \kern -13pt
                     \lower 3pt \hbox{$\sim$}}\ }

\def\lsim{\mathrel{\rlap{\lower4pt\hbox{\hskip1pt$\sim$}}
    \raise1pt\hbox{$<$}}}                % less than or approx. symbol
\def\gsim{\mathrel{\rlap{\lower4pt\hbox{\hskip1pt$\sim$}}
    \raise1pt\hbox{$>$}}}                % greater than or approx. symbol

\input epsf.tex
\def\desepsf(#1 width #2){\epsfxsize=#2 \epsfbox{#1}}

\newenvironment{tolerant}[1]{\par\tolerance=#1\relax}{ \par }
\usepackage{amsmath,bm}
\usepackage{lineno}
\usepackage{cite,./mcite}
\usepackage{tikz}
\usepackage[symbol]{footmisc}

\providecommand{\DOI}[1]{\href{http://dx.doi.org/#1}}

\begin{document}

\title{Humboldt Highway II - computer cluster on renewable energies}
\author[1]{Danyer~Perez~Adan~\orcidlink{0000-0003-3416-0726}}
\affil[1]{Rheinisch-Westf\"alische Technische Hochschule RWTH, Aachen, Germany}
\author[3]{Luis~Ignacio~Estevez~Banos\footnote{now at CTS Eventim AG \& Co., Hamburg}~\orcidlink{0000-0001-6195-3102}}
\author[2]{Tony~Cass~\orcidlink{0000-0001-9637-619X}}
\affil[2]{CERN, Geneva, Switzerland}
\author[3]{Bjoern~Felkers}
\affil[3]{Deutsches~Elektronen-Synchrotron DESY, Hamburg, Germany}
\author[4]{Fernando~Guzman~\orcidlink{0000-0002-7612-1488}}
\affil[4]{InSTEC, Universidad de La Habana, Havanna, Cuba}
\author[3]{Thomas~Hartmann~\orcidlink{0000-0003-4891-4584}}
\author[3]{Beate Heinemann~\orcidlink{0000-0002-1673-7926}}
\author[3]{Hannes~Jung~\orcidlink{0000-0002-2964-9845}}
\author[3]{Yves~Kemp~\orcidlink{0000-0001-9576-7850}}
\author[3]{Frank~Lehner}
\author[5]{J\"urgen~Nicklaus}
\affil[5]{Stefan Messer GmbH}
\author[4]{David~Gutierrez~Menendez}
\author[1]{Sandra~Consuegra~Rodriguez~\orcidlink{0000-0002-1383-1837}}
\author[4]{Cesar~Garcia~Trapaga~\orcidlink{ 0000-0001-5420-0540}}
\author[6]{Lidice~Vaillant~\orcidlink{0000-0003-1552-7449}}
\affil[6]{Universidad de La Habana, Havanna, Cuba}
\author[7]{Rodney~Walker~\orcidlink{0000-0001-8535-4809}}
\affil[7]{Ludwig~Maximilan~University, Munich, Germany}

\date{}
\begin{titlepage} 
\maketitle
\vspace*{-15cm}
\begin{flushright}
DESY-24-052
%\today
\end{flushright}
\vspace*{+14cm}

\begin{abstract}
In August 2023, IT experts and scientists came together for a workshop to discuss the possibilities of building a computer cluster fully on renewable energies, as a test-case at Havana University in Cuba.
The discussion covered the scientific needs for a computer cluster for particle physics at the InSTEC institute at Havana University, the possibilities to use solar energy, new developments in computing technologies, and computer cluster operation as well as operational needs for computing in particle physics. 

This computer cluster on renewable energies at the InSTEC institute is seen as a prototype for a large-scale computer cluster on renewable energies for scientific computing in the Caribbean, hosted in Cuba.

The project is called "Humboldt Highway", to remember Alexander von Humboldt's achievements in bringing cultures of the American and European continents closer together by exchange and travel. In this spirit, we propose a project that enables and intensifies the scientific exchange between research laboratories and universities in Europe and the Caribbean, in particular Cuba. 

\end{abstract} 
\end{titlepage}

\section{Introduction}
For many years scientists from DESY have cooperated with scientists from the InSTEC institute at Havana University in the field of Elementary Particle Physics. Students come nearly every year to attend the DESY summer student program in Hamburg, and many students from Havana have started and very successfully completed their PhD at DESY/Hamburg University.
 
In August 2023 more than 30 scientists and IT experts participated in the {\it Humboldt Highway II - computer cluster on renewable energies}~\cite{HumboldtHighwayII} to discuss possibilities for a full-scale computer cluster that runs fully on renewable energies. This workshop was funded with resources from the Hamburg Ambassador fund~\cite{HamburgAmbassadors} and was included in the program of the International Year of Basic Sciences for Sustainable Development (IYBSSD)~\cite{IYBSSD,IYBSSD-HumboldtHighwayII}.  
 
 \section{Computing Centers at CERN, DESY and InSTEC}
 Large-scale computing centers are operated at the big research centers, such as the international research center for particle physics, CERN in Geneva, and the national research center (with central service for universities), DESY in Hamburg. In the following, we briefly recap the main features of the different computing centers, with an emphasis on energy consumption.

 While renewable energies are not available all the time at the same intensity, strategies have to be developed for a stable operation of a computer cluster: 
 \begin{itemize}
 \item Operation on availability: the computer center is operated at full power only when renewable energies are available, otherwise the center is run in {\it sleep mode} and only critical tasks are performed. The {\it sleep mode} can be either achieved by switching off a fraction of the CPUs (as in a laptop, when put on sleep mode) or running CPUs at reduced frequency.
 \item Operation with battery systems: Huge battery systems are needed to run the computer cluster even during times when no power is delivered from renewable energy resources. At the price of enough battery storage, the computer cluster can be operated in a standard way.
 \end{itemize}
 
The energy consumption of a computing center depends on the number of CPUs and their power need: low-energy-consuming devices must be investigated (for example using ARM CPUs \footnote{as implemented in the latest Macbook M1, M2 series} instead of Intel/AMD x86 CPUs). Several WLCG sites are testing ARM CPUS with HEP workloads, and it is likely these will form pledgable resources in the near future. The operation of CPUs also produces heat, thus the cooling required is also to be investigated.
 
 \subsection{CERN}

CERN's existing data center houses computer and storage clusters with a total power consumption of 3MW and a new data center to support a further 4MW of computing equipment, growing to 8MW by 2028 and 12MW in the mid-2030s, is being commissioned. 
Nevertheless, CERN has long focused on the energy efficiency of its computing systems, a key point to consider for a solar-powered cluster, for two reasons.
Firstly, like all organizations nowadays, CERN wishes to minimize its environmental impact. More pragmatically, however, the capacity of its data centers is not limited by the availability of electricity to power the computing equipment but rather by the ability of cooling systems to remove the heat that is generated. 
The more efficient the computing systems are (in terms of Watts used per unit of compute power, data stored, or data transferred), the more useful work can be done within the limits of the cooling and ventilation systems.

Maximizing computing system energy efficiency obviously requires a focus on energy efficiency when procuring the computing hardware. Until recently, this has mostly meant concentrating on the efficiency of the power supplies (CERN requires these to have the industry standard 80 Plus Titanium rating) but, like many other sites, CERN is now investigating the energy efficiency of ARM-based systems for High Energy Physics workloads.
Once systems have been purchased, however, energy efficiency can be optimized by
\begin{itemize}
\item choosing the CPU operating frequency carefully as, in most cases, the highest operating frequency does not deliver the highest performance in terms of compute power per Watt, and
\item ensuring that systems run at an optimum load as power supply efficiency can often be poor when operated at less than 80\% of the rated load (although power supplies with the 80 Plus certification should perform well above 20\% of the rated load).
\end{itemize}

For the last point, the use of storage servers to deliver compute cycles should be considered as their data-serving task rarely requires a great fraction of their CPU capacity.
In the past, concern about the possible adverse impact on the primary task (whether simply from the additional load or from a malicious payload) was a reason to avoid such parasitic loads but today's container technology allows effective isolation of the compute workloads and servers are multi-cored so some of the cores can be reserved for the primary task.

Maximizing overall energy efficiency for a data center, though, requires attention to the cooling and ventilation system, not just the efficiency of the compute and storage clusters. The Power Usage Efficiency (PUE) of a data center is defined as the total power consumed by the center divided by the power delivered to the computing systems. Older data centers (such as the current data center) can have a PUE above 1.5, i.e. for every kW delivered to the compute systems, 500W is needed to power the cooling and ventilation system. 
Modern data centers aim for a PUE below 1.1 but even a 10\% overhead would be worth eliminating for a solar-powered data center. 

Fortunately, compute servers today do not need a precisely controlled environment with temperatures of 20\textdegree{}C or so but can operate with inlet temperatures of 25\textdegree{}C and above. In many cases (if the outside temperature is not too high), therefore, simply taking fresh air from the outside and ensuring that there is no way for exhaust air to return to warm the incoming air can be enough to provide adequate cooling, at least during a major portion of the year.

 \subsection{DESY}
 
The central IT department of DESY in Hamburg operates two large compute centers, with an approximate energy consumption of 1.3 MW. Power is drawn from two of the three main, independent power feeds of the Hamburg campus. UPS batteries are used to provide power in emergency situations for up to 60 minutes, as well as flattening out small disturbances in the power provisioning. Most of the racks are equipped with passive water-cooled rear doors. The heat captured this way is transferred to the DESY central cooling ring, where it is used for other purposes, e.g. assisting building heating.
The systems operated by the central IT department fall into several categories: Central infrastructure (e.g. mail, web, ...), scientific computing systems, and scientific storage systems. Different strategies apply to different systems w.r.t power efficiency.

The emphasis for the central infrastructure is clearly on availability and stable performance at all times. Past efficiency increase measures consisted mainly of server consolidation using virtualization technologies, which resulted in a reduced number of machines and thus power needed to operate the same services.

Scientific storage systems are the basis of all data analysis at DESY. Some storage systems are also integrated into the data acquisition of the experiments. Storage thus needs to be always available. Energy efficiency can be increased by carefully selecting the hardware setup, using an optimized hardware replacement strategy, and allowing the CPU frequency to react autonomously to changes in load from the compute infrastructure. Experiments can also define parts of their data as inactive. It can then be transferred to tape, where it is archived. Data storage on tape is very energy efficient, it needs nonetheless a good data management scheme.

Scientific computing systems are organized in interconnected farms of computers. Users put their compute jobs to a scheduling system, that controls dispatching to adequate computers. The same base-level energy efficiency measures apply to storage. More advanced energy efficiency measures include powering off whole nodes at times when they are not needed, or external factors such as energy availability or cost favor powering the nodes off. Integration with external decision-making is on the plan for the DESY scheduling systems, whereas an autonomous powering off based on needs is already implemented. Another advanced energy efficiency measure is reducing the frequency of the CPU at times when energy is scarce or expensive. Jobs will take longer and in the end, consume the same energy, but power consumption can be shifted without job losses to more advantageous times. The operating system provides some basic tools, for a concerted action, however, an integration of these tools into the scheduling system needs to be implemented. Again, this is planned for the DESY scheduling systems. It becomes clear that the scheduling system and scheduling, in general, plays an important role. We also plan to introduce an energy budget for computations. Users can decide whether they want to have results faster, which might induce a higher utilization of their energy budget since energy prices are higher. If users are OK to have results later, they are able to compute more, since their jobs get scheduled at times where energy prices are lower. At the moment, this is however only theoretical work, since DESY has a fixed energy price, that is independent of the production of green energy, or the consumption in the Hamburg area.

DESY also plans to look at architectures optimized for energy efficiency, e.g. ARM architecture. At the moment, however, many user codes are not yet available or validated for this architecture.

 \subsection{InSTEC}
InSTEC is currently a faculty of the University of Havana. Founded in 1987 as ISCTn (Institute for Nuclear Sciences and Technology), it was a university that attended the preparation of personnel for the Cuban nuclear program related to nuclear applications. Currently, it prepares students in 4 directions: Nuclear Physics, Radiochemistry, Engineering in Nuclear and Energy Technologies, and Meteorology. These fields require the handling of large data volumes, however, the required computer processing capacity is not available.

The current computer and storage devices at InSTEC consist of two clusters (abbreviated with C1 and C2 below) powered mostly by consumer hardware, i.e., Intel Core i3, i5, and i7 chips, non-ECC memory, and low-capacity HDD storage. Cluster interconnection uses a fiber-optic link at 1000 Mbps, from where cat6 ethernet is used to reach the nodes with speeds between 1000 Mbps and 100 Mbps.
Details of the compositions of C1 and C2 clusters are shown in Tab.~\ref{Tab1} and \ref{Tab2}.

\begin{table}[htp]
\caption{Details of the C1 cluster}
\begin{center}
\begin{tabular}{|c|c|c|c|c|c|}
\hline
\multicolumn{5} {|c|} {Computing Nodes} & Total \\ \hline
Type       & Core i3 & Core i5 & Core i7 & Xeon &   \\ \hline 
Number  &    1        &     4      &     2       &    1    &     8   \\ \hline 
Cores     &    2         &  16      &      8       &  16    & 42  \\ \hline 
Frequency (GHz) & 3.3 & 3.2 & 3.6 & 1.6 & \\ \hline
RAM (GB)   &    8  & 4   & 8  & 24 & 44 \\ \hline
HDD (GB)  & 500  & 1000 & 1000 & 4000 &  10500 \\ \hline
\end{tabular}
\end{center}
\label{Tab1}
\end{table}%

The C1 cluster is mainly dedicated to research in quantum mechanics applications in condensed matter physics and chemistry. The installed software packages are SIESTA, QMEXPRESSO, INTEL PARALLEL STUDIO 2015, DFTB+, MERCURY, GFORTRAN, GCC, SERPENT, ROOT, GEANT4, GROMACS, NewtonX, Anaconda3, MAMBA, GAMESS, and  WRF.

\begin{table}[htp]
\caption{Details of the C2 cluster}
\begin{center}
\begin{tabular}{|c|c|c|c|c|c|}
\hline
\multicolumn{5} {|c|} {Computing Nodes} & Total \\ \hline
Type       & Core i3 & Core i5 & Core i7 & Xeon &   \\ \hline 
Number  &    1        &     2      &     2       &    1    &     5   \\ \hline 
Cores     &    2         &  4      &     6       &  12    & 40  \\ \hline 
Frequency (GHz) & 3.3 & 3.2 & 3.6 & 2.4 & \\ \hline
RAM (GB)   &    8  & 8   & 8  & 16 & 56 \\ \hline
HDD (GB)  & 500  & 1000 & 1000 & 2500 &  7000 \\ \hline
\end{tabular}
\end{center}
\label{Tab2}
\end{table}%

The main tasks of the C2 cluster are Modeling, Simulation, and Data Analysis in Particle Physics, detector simulation, Nanomaterials, DFTB, and Medical Physics (Image processing), with dedicated application softwares CASCADE and PYTHIA.

Why two small clusters? Such a strategy was followed because the Institute has two independent power lines in two different circuits. With frequent power failures, the probability that both circuits simultaneously fail is smaller than the failure of one of them. In this way, the independent work of each cluster could be guaranteed. Though no automatic failure detection is in place, it is possible, with the aid of containerized solutions, to manually migrate critical tasks to the running cluster if needed.

Many challenges arise in maintaining operative computer resources at InSTEC due to multiple reasons, one of the most problematic is energy efficiency and availability. Despite continuous efforts to improve power efficiency across the institute, with a focus on heavy consumers like computing resources and cooling systems, it lacks modern international standards in terms of resiliency and efficiency, mostly due to outdated consumer hardware with less performance per watt and poor cooling solutions. 
Power outages are another big factor that adds uncertainty when managing the cluster operation and load, which is one of the main reasons for the involvement in the {\it Humboldt Highway II} project.

The geographical location of Cuba, with an average theoretical annual solar radiation value of the order of 5.4 kWh/m$^2$ 
\cite{SolarEnergyCuba,SolarEnergyCuba_WorldBank},  gives an average practical solar irradiance of about 4.5 kWh/m$^2$ 
which is considered high. Cuba with over 110,800 km$^2$ and an annual average of 330 sunny days presents excellent potential for the usage of solar energy. Considering the dependence of the country's electricity supply on fossil fuels with the existing and known limitations, it is tentative to bet on expanding the use of photovoltaic solar energy. It is also included in the National Economic and Social Development Plan until 2030 \cite{PlanNational}.

\begin{figure}[htbp]
\begin{center}
\vskip -3cm
\includegraphics[clip,width=0.98\textwidth]{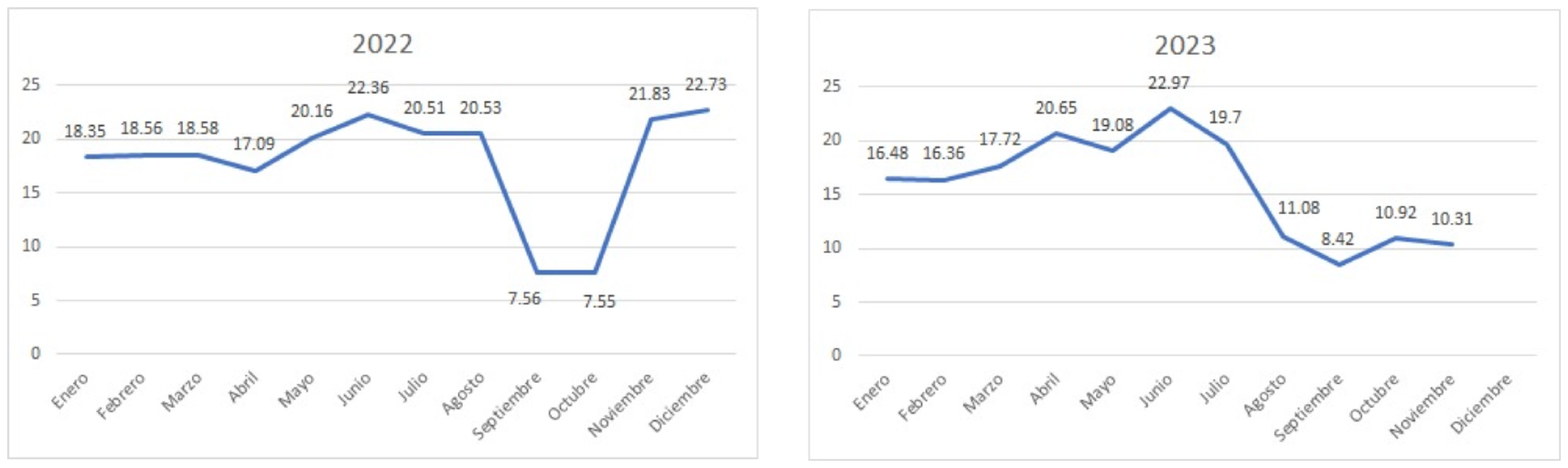}
\vskip -3cm
\caption{Monthly power consumption (MWh) in the last two years}
\label{Fig1}
\end{center}
\end{figure}

Figure~\ref{Fig1} shows the institute's yearly power consumption. There are no significant variations from one year to the other. As expected, the maximum power consumption corresponds to the period between April and August due to the high temperatures and the need to use air conditioning for the equipment. On the other hand, it is this period in which the solar irradiance is highest.

\subsubsection{Computing Center Operation: Load-shedding}

As the proportion of intermittent renewable electricity increases,  the ability to vary consumption dynamically becomes increasingly important. Load-shedding is the on-demand reduction of electrical consumption, in order to match the available production. Such reductions happen on various timescales, from those planned days in advance, based on usage patterns and weather forecasting, to the very rapid modulation used to stabilize the grid frequency. 
High-throughput computing has the potential to easily and quickly reduce power consumption on demand. The fastest response is achieved by lowering the CPU frequency, which can reduce the power consumption by 50\% instantly.  The running workloads of course go slower, but the action is reversible and can be done frequently. This is ideal for following hourly fluctuations, typical of solar energy and peak demand.
Longer periods of low load can be achieved by draining and powering down nodes. This could be useful for interactive nodes that are not needed outside working hours. 
Short periods can be covered by a standard solar battery system, where the grid load is replaced by battery power. A carefully sized photovoltaic and battery system has the potential to cover all the power needs of a Cuban compute cluster for most of the year.  Even when grid power is required, the batteries protect against frequent outages, for example, during hurricane season. In this scenario, the load-shedding techniques provide the flexibility to operate with a smaller battery capacity.

In Europe, load-shedding techniques are interesting due to the varying carbon intensity and price of grid electricity, which are ideally strongly correlated. Reducing consumption when the price is high, allows us to run the older less efficient fraction of available hardware when the price is low. In this way, we can do the same work for less money and lower carbon footprint \footnote{Assuming the price is correlated with the carbon intensity, so running less efficient hardware when the price and the carbon intensity are low.}, 
as well as contributing to grid stability. The tumbling price of batteries means they may also play a role.

\section{Energy sources}
The options for the use of renewable energies depend on the geographic location of the research center: in northern Europe, energy is mainly produced by wind, while in southern regions, solar energy will be the main source. Both resources have in common, that they are not available at full strength all the time, while the availability of solar energy and wind energy may be significantly different, therefore a proper system has to be flexible enough to cover the different cases.

Feeding computing clusters with renewable energies is an effective way to reduce their environmental impact and promote sustainability. One approach is to install solar modules or wind turbines near the data center to generate clean electricity. These renewable sources can be integrated into the power grid, ensuring a continuous and reliable energy supply. Excess energy can even be stored in batteries for use during periods of low renewable generation.
 
Another strategy involves sourcing renewable energy from off-site locations, such as purchasing renewable energy credits (RECs) or signing power purchase agreements (PPAs) with renewable energy providers. This allows data centers to indirectly support renewable energy generation while still receiving stable power from the grid.
 
To maximize efficiency, computing clusters can employ various energy-saving techniques. These include optimizing workload distribution, using energy-efficient hardware components, implementing advanced cooling systems, and adopting virtualization technologies to consolidate workloads.
 
Overall, feeding computing clusters with renewable energies not only reduces their carbon footprint but also contributes to the global transition towards a cleaner and more sustainable energy future.

The development of systems to plan and vary the power consumption, in response to local production, battery SOC or grid price, is required in both the Cuban and European scenarios, albeit for slightly different reasons.

Since wind turbines in urban landscapes are more complex to exploit due to turbulence effects, solar photovoltaic is the main technology dominating the penetration of renewable energy solutions in urbanized areas.

While there are numerous advantages to powering computing clusters with renewable energies, there are additional challenges to be considered: 
\begin{itemize}
\item renewable energy sources such as solar and wind are subject to natural variations and intermittency. The availability of sunlight and wind can fluctuate, leading to inconsistent power generation. This intermittency is the main challenge in ensuring a stable and reliable power supply to computing clusters.
\item the grid integration of renewable energy sources into the existing power grid can be complex. The grid infrastructure may need to be upgraded or modified to accommodate the intermittent nature of renewable energy generation. Additionally, issues such as grid congestion and transmission limitations can impact the efficient integration of renewable energy into the grid.
\item while the cost of renewable energy technologies has been decreasing over time, it can still be relatively expensive for an initial investment, in particular in developing countries. The initial investment required for setting up renewable energy infrastructure, such as solar panels or wind turbines, can be significant.
\item space requirement should also be considered. Generating large amounts of renewable energy often require substantial land or space for installing solar panels, wind turbines, or other infrastructure. In densely populated areas, finding suitable locations for renewable energy installations can be challenging.
\end{itemize}
Many of these issues can be mitigated or overcome with advancements in technology, grid infrastructure improvements, and better energy management strategies.

In the following, we discuss in detail solar energy, and we leave the usage of wind energy to another report.

\subsection{Photovoltaic Systems}

Different ways for installing photovoltaic systems to be used to power computing clusters are possible:
\begin{itemize}
\item Roof-top solar systems: These are the most common types of photovoltaic systems. Solar panels are installed on the roof-tops of buildings, including data centers, to capture sunlight and convert it into electricity.
For the roof-top solar systems, there are two configuration options:
\subitem{-} grid-tied, meaning they are connected to the local power grid, with or without batteries as a backup system for cut-off and night consumption or
\subitem{-} off-grid, where excess energy is stored in batteries for later use.
\item Solar farms: Solar farms consist of large-scale arrays of solar panels installed on open land or in dedicated areas. They generate a significant amount of electricity and can be connected to the power grid to supply renewable energy to computing clusters. This option would not ensure the functioning of the computing cluster during a power cut unless a backup battery system is installed. By investing in solar farms, even if located far away from the consumption point, the facility can claim to be renewable.
\item Building-Integrated PhotoVoltaics (BIPV): BIPV systems are integrated directly into the architecture of buildings, such as solar windows, solar roof tiles, or solar facades. These systems combine the functionality of traditional building materials with solar power generation, allowing computing clusters to be powered by solar energy without the need for separate solar panels. This option is interesting in cases in which the computing cluster is being implemented as a functional part (i.e. inside) of a building project.
These different types of photovoltaic systems provide flexibility in harnessing solar energy to power computing clusters, catering to various installation requirements and locations.
\end{itemize}

There are several considerations to keep in mind when designing a photovoltaic system for a computing cluster. Here are some important factors to consider:
\begin{itemize}
\item[1] Energy Demand: It's crucial to accurately estimate the energy demand of the computing cluster. This includes considering the power requirements of servers, networking equipment, cooling systems, and other components. Understanding the energy consumption patterns and load profiles will help determine the size and capacity of the photovoltaic system. In this point, the energy efficiency analysis is crucial.
\item[2] System Sizing: The photovoltaic system should be appropriately sized to meet the computing cluster's energy demand. This involves considering factors like the available roof or land area for solar panels, the solar radiance of the location, and the efficiency of the photovoltaic modules. Sizing the system correctly ensures that it can generate sufficient electricity to power the cluster.
\item[3] Storage and Backup: To address intermittency and ensure a stable power supply, it may be necessary to incorporate energy storage systems, such as batteries, into the photovoltaic system. This allows excess energy generated during peak sunlight hours to be stored and used during periods of low or no solar energy generation.
\item[4] Grid Integration: Depending on the computing cluster's energy needs, it may be necessary to integrate the photovoltaic system with the local power grid. This allows for power exchange between the cluster and the grid, ensuring a reliable supply of electricity when solar generation is insufficient or excess energy can be exported back to the grid.
\item[5] Scalability and Future Expansion: Consider the potential for future growth and expansion of the computing cluster. The photovoltaic system should be designed with scalability in mind, allowing for the addition of more solar panels or modules as the cluster's energy demand increases over time.
\item[6] Maintenance and Monitoring: Implementing a robust maintenance and monitoring plan is vital for the long-term performance and reliability of the photovoltaic system. Regular inspections, cleaning of solar panels, and monitoring of system performance will help identify and address any issues promptly.
\end{itemize}
By considering these factors during the design phase, a well-designed photovoltaic system can effectively power a computing cluster with renewable energy, ensuring sustainability and reducing environmental impact.

\section{The Humboldt Highway project}
In 2013  scientific exchange between the InSTEC institute of Havana University and DESY started, first with summer students who joined the DESY summer-student program. Since then, many students visited DESY for the summer-student program, and several started a PhD at DESY and the University of Hamburg. Since 2017 senior scientists from InSTEC have visited DESY regularly. 

While there is significant progress when scientists meet in person, the progress is slowed down when they are back in Cuba, since the internet connection and computing facilities are limited.
In order to allow for more continuous scientific exchange over the whole year, as well as to allow more students to participate in international scientific discussions, DESY and Havana University initiated the "Humboldt Highway" project which aims to establish and operate reliable and stable internet connection. 

The goal of the Humboldt Highway project is to launch an intense scientific exchange between DESY/University Hamburg and Havana as well as to initiate scientific schools to  foster and complement the education of  students in Cuba. A phenomenology group in the area of Quantum Chromo Dynamics at the InSTEC institute of Havana University could be created to allow participation in particle physics phenomenology projects at DESY as well as to participate in particle physics experiments at the Large Hadron Collider (LHC) at CERN in Geneva. 

Internet connection is a crucial element in today's scientific exchange. During the pandemic, it became very obvious that without the possibility to participate and to hold video conferences, the digital divide would even further decouple countries like Cuba from fruitful  discussion and development. The Humboldt Highway project is also meant to overcome the isolation of Cuba as a consequence of the embargo against Cuba.
It will allow the scientific community as well as students in Cuba to take part in international collaborations.  
The Humboldt Highway project aims against a brain drain from Cuba to Europe and intends to contribute to building and developing the scientific structures inside Cuba.

The Humboldt Highway project is fully in line with the bilateral relations between Germany and Cuba~\cite{BilateralCubaGermany} as well as with the relations of the European Union with Cuba \cite{EU-Cuba}. 

The internet connection and computing facilities in Cuba are the essential bottlenecks.  Even the satellite system operated from StarLink is not accessible from Cuba \footnote{Email exchange with StarLink office on 2023-06-01}.  Moreover, connection to ZOOM is disabled \cite{ZOOM-restrictions}, and connections are only possible via third parties. 

Further severe complications are export restrictions for material and devices to be sent to Cuba (for an overview see \cite{ExportRestrictions}). In addition to the general export control regulations under EU law, the comprehensive restrictions under US (re-)export control law in particular must be observed and complied with. US goods (commodities, software, and technology) are subject to the EAR (US Export Administration Regulations) and therefore subject to the embargo. Hence, it is not allowed to export US goods to Cuba. This also applies to so-called "foreign-made products" (e.g. Made in Germany) that contain more than 10 percent US goods in terms of value~\cite{export-control1}. It is therefore advisable to oblige the supplier to provide the relevant information. A plausibility check of the information received should then be carried out. In case of doubt, the product should not be delivered to Cuba or only with official approval (export license from US authorities)~\cite{export-control2}.   

The general terms and conditions of suppliers/manufacturers of the products must also be carefully checked and complied with. Some suppliers specify that their products in general may not be (re-)exported to certain countries. Moreover, there is the practical problem that some transport companies and logistics service providers do not deliver goods to Cuba at all.

Therefore to help advance scientific development in  Cuba, the only practical and short-term option is to construct and operate computing facilities on site. If it is intended in this context, then deliveries of material, devices, and software to Cuba must be legally compliant with all applicable customs and export control regulations. Close cooperation between project management, purchasing, export control department, and logistics provider (agent) is recommended throughout the process. 

Since delivery of energy is still much affected by the blockade against Cuba, and power cuts are happening regularly, a computer cluster should be operated independently from the public electricity circuits to ensure energy supply for critical infrastructure, like hospitals, etc.  The geographical location of Cuba makes the use of solar energy obvious.

\subsection{A first step: a computer cluster on renewables for particle physics at InSTEC}

The {\it Humboldt Highway II} project to construct a computer cluster operating on renewable energies in Cuba is mainly targeted at students of the University, taking into account the role they will play in the era of information technology and big data science, their needs to become full members of international cooperation, to play a role at the frontier of sciences, and their role in multiplying knowledge.

As a first step, with a timescale of 1--2 years, a small computer cluster fully on renewable energies could be constructed.  To achieve this,  
 the photo-voltaic (PV) system needs to be designed including charging batteries. A small computer cluster could be perhaps built with laptops (using their own batteries) running ARM CPUs, allowing easier handling of the sleep-mode option. For combining the computer nodes into a cluster, involvement from IT experts would be  needed.
 
With such a small cluster, a limited scientific program in terms of participation in phenomenology projects in particle physics can be achieved.
\begin{itemize}
\item To contribute to preparing young people to work with computing clusters.
\item To demonstrate that such a solution, based on photovoltaic energy supply, is a practical way to contribute to sustainable development.
\item A small cluster, as considered in the {\it Humboldt Highway II} project, will be a practical example to be considered for other areas of the Cuban development programs. One can mention, as one example, the possibilities that it shows the way in which it could be introduced in the medical care program which is structured as a system of small distributed Clinics for primary medical assistance.
\end{itemize}

\subsection{Thinking bigger: a computer cluster on renewables for Cuba}

With the experience of a small-scale computer cluster, a large-scale cluster can be designed and constructed. For this, new developments in the area of PV systems, battery storage, and IT technology as well as computing algorithms and system integration are required. The investigations performed here can serve as a testbed for a large-scale cluster in different countries and regions.

\subsection{Thinking big: a computer cluster on renewables for the Caribbean}
The large-scale computer cluster in Cuba can be further extended to serve as a Caribbean hub, bringing together the different countries in the Caribbean to use the computing resources of an international computer hub in Cuba, under the umbrella of UNESCO. Such a hub would serve the Caribbean and international research communities with a vision as open, collaborative platforms for basic and applied science addressing also global challenges, for instance, the rising computing needs in region-specific models of climate change, modeling and forecasts of extreme weather events, or health and life science applications. 

Such a center would serve the same spirit as was for creating CERN~\cite{CERNmission} in 1954 bringing scientists of countries together, who were at war against each other a few years before, and similarly the SESAME~\cite{SESAME-home,SESAME-unesco} project in the Middle East or the SEEIST~\cite{SEEIST-home} in South-East Europe. 

\subsection{Funding and Partnership Considerations}
In general, digital partnerships and cooperation in information and communication technologies are indispensable for accelerating progress towards the Sustainable Development Goals (SDGs) of the United Nations and for unlocking the transformative potential of the digital economy. Hence, the roadmap towards a Caribbean computing hub could be a partner of the International Decade of Sciences for Sustainable Development~\cite{IYBSSD}  and a model for other regions to participate in this decade and maybe one day to create and participate in an International Organization for Sustainability Sciences.

However, dedicated research and development are needed to further advance the idea of a scalable, modular, energy-efficient computer cluster based on renewable energies. Funding could come from grants from Horizon Europe, e.g. through a European Research Grant ERC covering the three pillars: renewable energy supply and battery storage, computer cluster operation depending on availability of energy and optimization of computer algorithms, or through the DIGITAL Europe program, a new EU funding program focused on bringing digital technology to businesses, societies, and public administrations including digital partnerships with the Global South.
Moreover, there are UN programs such as the UN agency for Information and Communication Technology ICT which offers its own Development Funds (ICT-DF) contributing to sustainable development through the co-financing of ICT national, regional, and global development projects.

The vision of a scalable, renewable, and energy-efficient computer cluster in the Caribbean can also become an attractive element in the strategic framework of the EU-CELAC~\cite{EU-CLEAC-2023} (Community of Latin American and Caribbean States) bilateral platforms. Just recently, a "Digital Alliance" was launched in March 2023 with the ambition to join forces for an inclusive and human-centric digital transformation in both regions and to develop bi-regional dialogue and cooperation across the full spectrum of digital issues. Large synergies can also be expected from the EU-Africa Innovation agenda.

\section{Conclusion}

In August 2023, IT experts and scientists came together for a workshop to discuss the possibilities of building a computer cluster fully on renewable energies, for a testbed at Havana University in Cuba.

During the discussion, the idea of a full-scale computer cluster was created which could serve the Caribbean area for high-scale computing in parallel as a project against the scientific and digital divide, bringing scientists from different countries together and bridging over any blockade.

Globally, efforts are needed to develop digital and equitable partnerships with developing countries to accelerate progress on the Sustainable Development Goals. CERN, DESY, and other research labs and institutes could play here a special role as catalysts and help with their international networks and expertise to move forward with a roadmap for a scalable and energy-efficient computer cluster based on renewables in Cuba and later in the Caribbean. Such a roadmap would also include pilot projects and seed funding capabilities as well as the development of robust, transparent, and international governance on open science and open data policies. 

\vskip 0.5 cm 
\begin{tolerant}{8000}
\noindent 
{\bf Acknowledgments.} 
We are very grateful to {\it Hamburg Ambassadors} for the encouragement and financial support for this workshop.
We thank DESY for hospitality and support with infrastructure for organizing the Humboldt Highway II workshop..
 \end{tolerant}

\bibliographystyle{mybibstyle-new.bst}
\raggedright  
\input{writeup-2024-04-09.bbl}

\end{document}

%% file: writeup-2024-04-09.bbl
\providecommand{\href}[2]{#2}\begingroup\raggedright\endgroup